# Direct visualization and effects of atomic-scale defects on the optoelectronic properties of hexagonal boron nitride

*Pablo Ares[1,2]\*, Hernán Santos[3], Snežana Lazić[2,4,5], Carlos Gibaja[2,6,7], Iñigo Torres[2,6,7], Sergio Pinilla Yanguas[2,4,5], Julio Gómez-Herrero[1,2], Herko P. van der Meulen[4,5], Pablo García-González[2,8,9] and Félix Zamora[2,6,7]\**

[1] Departamento de Física de la Materia Condensada, Universidad Autónoma de Madrid, 28049 Madrid, Spain

[2] Condensed Matter Physics Center (IFIMAC), Universidad Autónoma de Madrid, 28049 Madrid, Spain

[3] Departamento de Matemática Aplicada, Ciencia e Ingeniería de Materiales y Tecnología Electrónica, ESCET, Universidad Rey Juan Carlos, C/ Tulipán s/n, 28933 Móstoles, Madrid, Spain

[4] Departamento de Física de Materiales, Universidad Autónoma de Madrid, 28049 Madrid, Spain

[5] Instituto Universitario de Ciencia de Materiales "Nicolás Cabrera" (INC), Universidad Autónoma de Madrid, 28049 Madrid, Spain

[6] Departamento de Química Inorgánica, Universidad Autónoma de Madrid, 28049, Madrid, Spain

[7] Institute for Advanced Research in Chemical Sciences (IAdChem), Universidad Autónoma de Madrid, 28049, Madrid, Spain

[8] Departamento de Física Teórica de la Materia Condensada, Universidad Autónoma de Madrid, 28049 Madrid, Spain

[9] European Theoretical Spectroscopy Facility (ETSF)

\* E-mail: pablo.ares@uam.es; felix.zamora@uam.es

Hexagonal boron nitride (hBN) is attracting a lot of attention in the last years, thanks to its many remarkable properties. These include the presence of single-photon emitters with superior optical properties, which make it an ideal candidate for a plethora of photonic technologies. However, despite the large number of experimental results and theoretical calculations, the structure of the defects responsible for the observed emission is still under debate. In this work, we visualize individual atomic-scale defects in hBN with atomic force microscopy under ambient conditions and observe multiple narrow emission lines from optically stable emitters. This direct observation of the structure of the defects combined with density functional theory calculations of their band structures and electronic properties allows us to associate the existence of several single-photon transitions to the observed defects. Our work sheds light on the origin of single-photon emission in hBN that is important for the understanding and tunability of high-quality emitters in optoelectronics and quantum technologies.

**1. Introduction**

Hexagonal boron nitride (hBN) is a layered material with a unique combination of remarkable properties, including exceptional strength,[1] oxidation resistance at high temperatures,[2] piezoelectricity in its single layer form,[3] it has become a key component for a majority of the two-dimensional technologies[4] and presents novel photonic functionalities,[5] to name but a few. Among the latter, we can find multiple recent research on the bright and stable single-photon emission at room temperature from atomic-scale defects in hBN[6-21] because of its potential for quantum information processing. The role of atomic-scale defects in solids is essential for their mechanical, electronic or optical properties and, despite the many experimental studies and theoretical calculations to unveil the nature of the defects causing these light emissions,[22-29] the possible crystalline structure of these defects is not completely clear. This is in part due to the shortcomings for direct visualization of the atomic-scale defects present in hBN. State-of-the-art aberration-corrected Transmission electron microscopy (TEM) techniques allow discriminating the boron and nitrogen atoms of hBN and have been widely used to image hBN atomic defects,[7, 30-38] although the elemental assignment of single atoms with small atomic numbers presents some limitations. The use of scanning tunneling microscopy (STM), one of the most extended tools for imaging atomic point defects in conductors, semiconductors and ultrathin films, remains elusive for insulators in general, and in hBN (a wide bandgap insulator) in particular, as it needs to establish a tunnel current between the probing tip and the sample. Intrinsic charged BN defects have been indirectly visualized and manipulated with STM by capping a BN crystal with a monolayer of graphene.[39] TEM and STM are typically employed under vacuum conditions and, to the best of our knowledge, no direct visualization of hBN atomic defects in ambient conditions has been reported. In this work, we use atomic force microscopy (AFM) to visualize atomic-scale defects in hBN under ambient conditions and we observe photoluminescence spectra with multiple narrow emission lines originated from bright and optically stable emitters. We then obtain the electronic and optical properties of the observed point defects within the framework of Kohn-Sham density functional theory. Our results suggest the existence of several single-photon transitions associated with the different defect structures imaged, clarifying the role of atomic-scale defects on hBN flakes on their optical properties.

**2. Results and discussion**

In this work, we started from commercially available hBN powder in the form of grains with an average diameter of ~ 1 μm (Sigma-Aldrich), which we thermally expanded as reported elsewhere[12] to obtain few-layer hBN with single-photon emitters. For an initial optical and atomic force microscopy characterization (**Figure 1**), we suspended the expanded hBN in a 2-Propanol/$H_2O$ dispersion and casted on $SiO_2$/Si substrates and dried after 15 minutes under an Ar flow (see Methods for details).

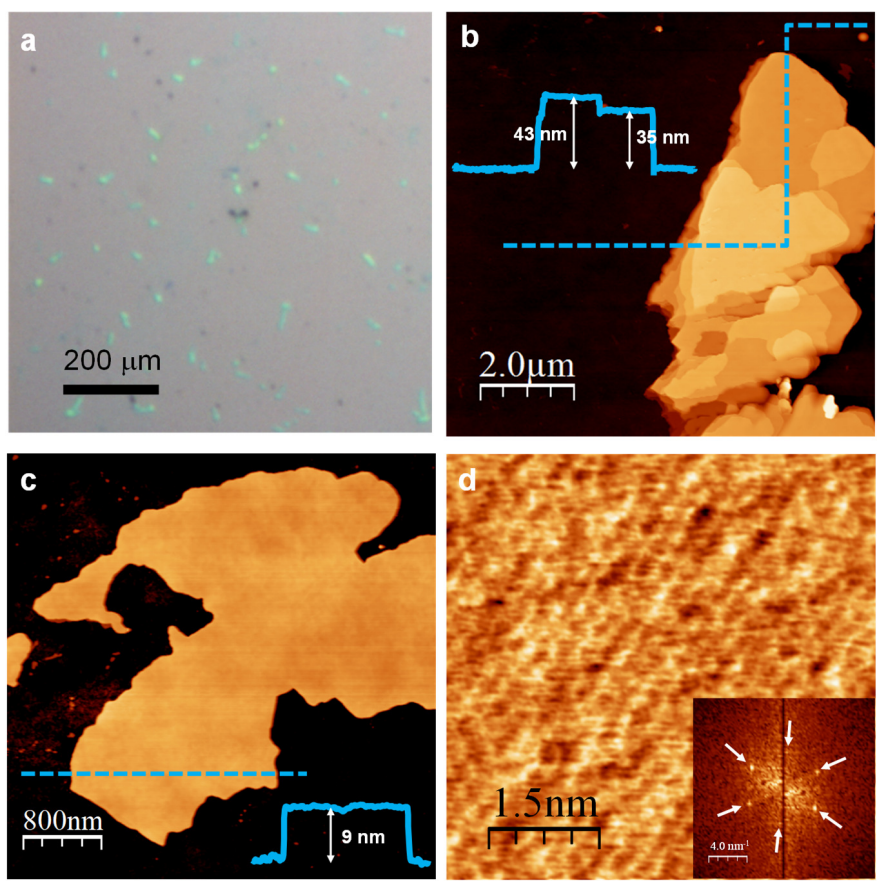

**Figure 1.** Initial characterization of hBN. a) Optical microscopy image showing few-layer hBN flakes drop-casted on SiO$_2$/Si. A homogeneous dispersion of flakes can be observed. b) AFM topographical image of a flake presenting different height terraces. Notice the angles observed in the edges of the terraces, mainly multiples of 60º, as expected for a hexagonal structure. The inset corresponds to the height profile along the dashed line. c) AFM topographical image of another flake, in this case with a homogeneous height of ~ 9 nm. d) High resolution AFM imaging on an hBN flake where the atomic lattice is clearly visible. The image has been high and low pass filtered to optimize its quality (see Methods for details). The inset shows its Fast Fourier Transform (FFT), where the arrows are pointing to the main peaks displaying the hexagonal arrangement of the lattice.

Figure 1a shows an optical microscopy image where few-layer hBN flakes are clearly visible homogeneously distributed on the surface. Micron-scale AFM characterization (Figure 1b and 1c) shows few-layer hBN flakes of several thicknesses, from tens of nm down to just a few nm. Contact mode high resolution AFM imaging on these flakes allowed us to clearly distinguish the hexagonal atomic arrangement of hBN (Figure 1d) with a lattice parameter of ~ 2.5 Å, assessing the high quality of our few-layer hBN flakes. Scanning different areas of the flakes at this resolution led to the visualization of possible defects within the atomic lattice. We chose contact mode because i) it allows avoiding possible artifacts in height measurements compared to dynamic modes,[40] and ii) it can achieve "pseudo-atomic resolution" or "lattice resolution" topographic images in hBN and other layered materials,[41-43] and it can as well achieve what it is called "true atomic resolution". Although the mechanisms governing this high resolution are not entirely understood, it is characterized by the possibility of distinguishing atomic defects on the surface.[43-46] **Figure 2a** shows an 11×11 nm$^2$ area at a high resolution where a defect is highlighted within a dashed blue circle. Figure 2b is a zoom in of this same defect, where we have superimposed the atomic lattice of hBN, and Figure 2c and 2d correspond to other defects we visualized. From our data, we cannot identify the nature of the missing atoms,[47, 48] which is out of the scope of this work. However, based on the high resolution TEM observations of Jin *et al.*[34], we can overlay a tentative most probable configuration of the defect in Figure 2b, which results in a 3$V_B$ + $V_N$ (missing three boron and one nitrogen atoms). Following the same procedure, in Figure 2c we can distinguish $V_N$, $V_B$ and $V_N$ +$S_{B \to N}$ vacancies, and the defect in Figure 2d is compatible with $V_N$ +$S_{B \to N}$. We observed more defects, including single-, two-, four- and 9-atom vacancies (see Figure S1 in the Supporting Information). Please note that, since we cannot individually resolve the nature of each of the atoms, the $V_B$ could be $V_N$ and *vice versa*, the $V_N$ +$S_{B \to N}$ could be $V_B$ +$S_{N \to B}$ and the four-atom 3$V_B$ + $V_N$ defects could be as well 3$V_N$ + $V_B$ vacancies. It is important to remark that all these possible defects in hBN have been detected by different techniques[33, 34, 36, 39, 49-52]. Additionally, our observations were stable over time in ambient conditions: we were able to visualize the defects without significant changes for several consecutive scans (Figure S2 and S3) and at different scanning angles (Figure S4).

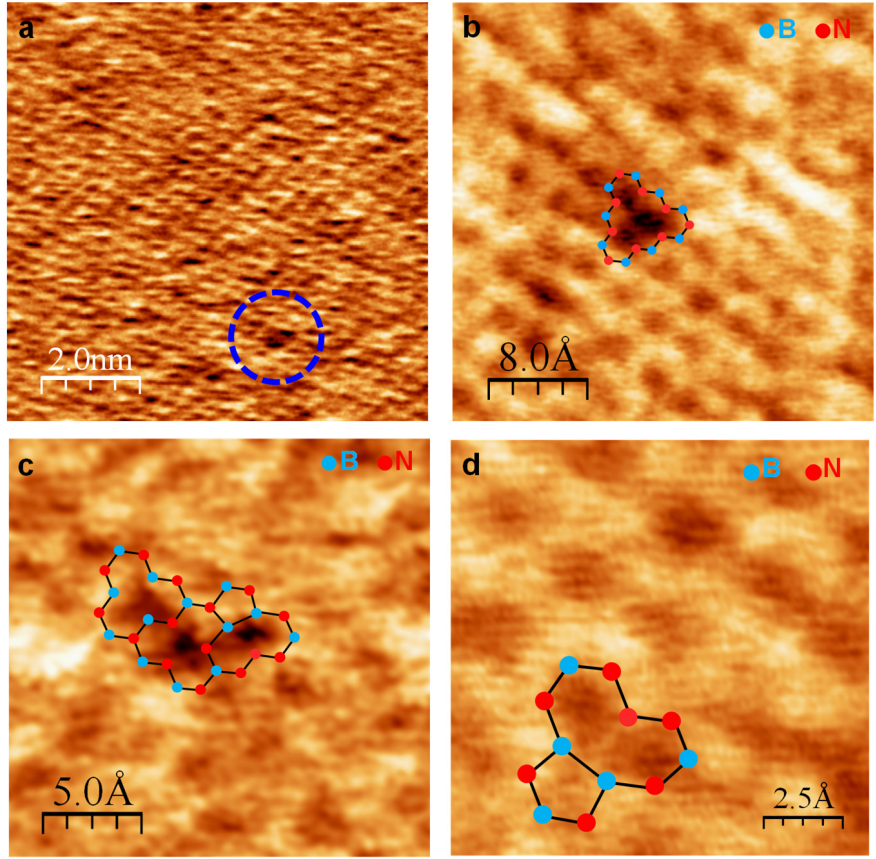

**Figure 2.** High resolution AFM imaging on hBN flakes. a) 11×11 nm$^2$ area showing the atomic lattice of hBN. The dashed blue circle depicts an atomic defect. b) Zoomed image of the defect in a), where a tentative schematic of the hBN atomic lattice around the defect is overlaid to help identify the defect, which is compatible with a 3$V_B$ + $V_N$ vacancy. c) Tentative $V_N$, $V_B$ and $V_N$ +$S_{B\rightarrow N}$. d) Tentative $V_N$ +$S_{B\rightarrow N}$. The images have been high and low pass filtered to optimize their quality and highlight the defects (see Methods for details).

Next, we carried out micro-photoluminescence characterization of our material. To this end, we placed a small amount of grains consisting of a collection of thermally expanded hBN flakes on a SiO$_2$/Si substrate and mounted the samples in a liquid Helium closed loop cooled cryostat (see Methods for details). We observed emission of the defect centers in the spectra taken at room temperature (**Figure 3a**) where the emission peaks were found to be strongly localized, within the precision of our laser spot of 1.5 μm. At low temperature (Figure 3b), these peaks narrow, and many more defects were emitting in this case.

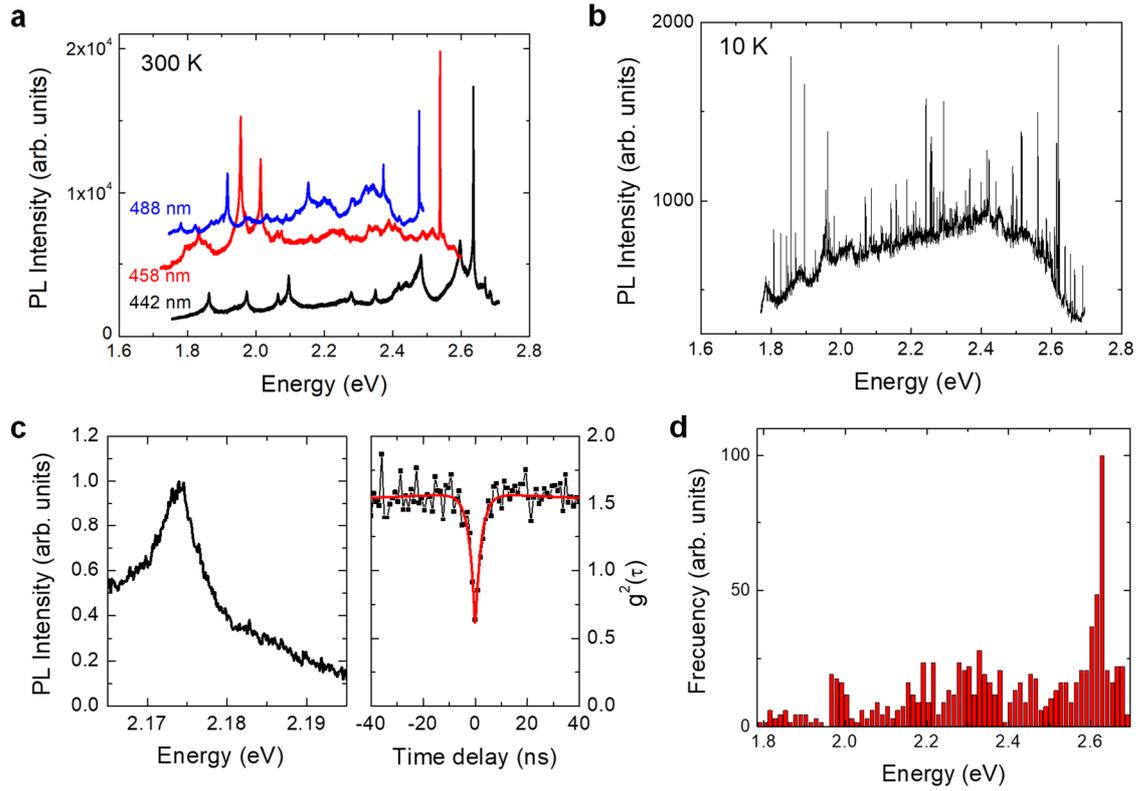

**Figure 3.** Photoluminescence results. a) Photoluminescence at room temperature, with different excitation wavelengths ($\lambda_{exc}$ = 488 (blue line), 458 (red line), and 442 (black line) nm, offset for clarity), on different grains of hBN flakes. b) Photoluminescence at 10 K, with 442 nm excitation, showing a large number of very narrow defect emissions on top of a continuous background. c) Room temperature defect emission and correlation function $g^2$ showing single-photon emission by antibunching, see ref. [12] for details. d) Histogram of the line positions of 322 narrow emission lines for several grains of hBN flakes, the maximum occurrence at 2.63 eV corresponds to the energy of 1LO phonon below the excitation energy.

Photon correlation measurements at room temperature and at 10 K have shown that they are single-photon emitters, as has been observed by a number of authors,[9, 19] and is illustrated in Figure 3c for a room temperature emission. The emission of these centers is linearly polarized, with a random polarization orientation.[12] The temporal evolution of the photoluminescence signal at low temperature (see Figure S5) shows that some centers were stable in time while others showed spectral diffusion. We also observed the effect of blinking. The low temperature emission shows narrow peaks in a wide energy range, between 1.7 to 2.7 eV, where the upper limit is only determined by the laser excitation (the data were taken with an excitation at 442 nm). To investigate the occurrence of these peaks, the statistics of the line positions is represented in a histogram in Figure 3d using the data of 322 lines. We observe a maximum at a value of 2.63 eV. Unfortunately, this does not give information about the nature of the defects. The maximum can be explained simply by a resonance with the excitation energy, it occurs at an energy distance corresponding to 1 LO phonon (170 meV) below the excitation energy of 2.805 eV, enhancing the emissions around that value. The same can be observed in results reported in the literature,[9, 11] although these were not identified as such. A direct relation between defects observed in photoluminescence and AFM measurements cannot be made as they collect data over largely different sized areas.

In order to gain insight into the origin of the observed emission peaks, we studied the electronic and optical properties of the observed defects within the framework of Kohn-Sham density functional theory (KS-DFT)[53, 54] (see Methods for details). The signature on the optical properties of simple $V_N$, $V_B$ monovacancies and of the highly unstable $V_{BN}$ divacancy has been already analyzed using KS-DFT[7, 25, 38] and ab-initio many-body perturbation theory (MBPT)[23] calculations. Although there are some discrepancies between different KS-DFT results, neither KS-DFT nor MBPT predicts the existence of a transition between localized defect states within the gap for those defects. It is worth noticing that the KS-DFT approximated by the GGA functional absorption spectra for the $V_N$ monovacancy exhibits a well-defined spectral feature peak around 1.9 eV, but it is due to transitions from an occupied localized defect state to extended states near the edge of the conduction band. Exchange and dynamical screening processes described under the GW approximation blueshift such a feature to 4.1 eV, but the eventual inclusion of excitonic effects (electron-hole interactions) leads to a definite energy of 2.7 eV,[23] which provides a rough estimation of the expected errors in the KS-GGA mean field approximation.

Using KS-DFT calculations, Tran *et al.* proposed that localized states associated to a complex defect made up by a $V_N$ vacancy plus an N atom occupying one of the closest B sites ($V_N + S_{B \rightarrow N}$) may be responsible for single-photon emission.[19] Our KS-GGA result does confirm that finding (**Figure 4**): the absorption spectrum for this defect exhibits a well-defined transition at 1.95 eV involving two localized states: the HOMO level (labeled as A in Figure 4) and an unoccupied one (labeled as B), which is 0.8 eV below the conduction band. The spatial distribution of the A and B states clearly indicates that the A → B transition is centered in the substitutional site, that is, around the N atom. Moreover, this transition is quite stable against small distortions of the defect geometry and against passivation of the N and B dangling bonds (see Figure S6 and S7 in the Supporting Information). In addition, there is a continuum of transitions starting at 2.7 eV that mainly comprise excitations from the HOMO level to extended states in the conduction band, and from valence band states to the LUMO level. These two series of transitions prevent clear identification in the spectrum of further excitation processes between localized states.

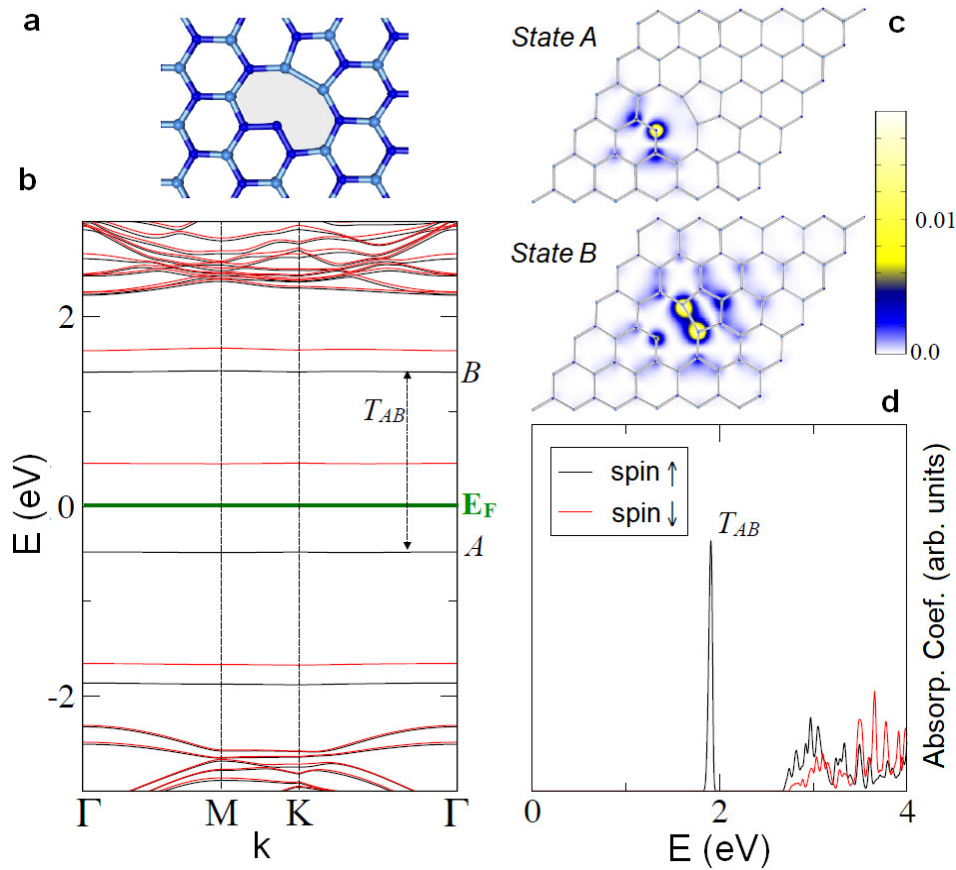

**Figure 4.** Electron and optical properties of the anti-site $V_N + S_{B \to N}$ defect. a) Geometry of the defect. b) Spin-resolved KS band structure. $T_{AB}$ is the optically allowed transition between the occupied defect states A and the unoccupied one B. c) Representation of the projected LDOS of states A and B. d) Absorption spectrum corresponding to perpendicularly incident unpolarized light as obtained from the sum of the *xx* and *yy* components of the independent-particle dielectric tensor.

The abundance of high resolution images that can be attributed to the triangular shapes $3V_N + V_B$ and/or $3V_B + V_N$ defects opens the question of whether these structures can also accommodate transitions between localized defect states. The optical absorption spectrum for the $3V_N + V_B$ shows very weak features at 3 eV and 3.75 eV that might correspond to transitions between defect states (Figure S8). However, the intensity and frequency of those transitions are very sensitive to the structural details of the defect. Furthermore, many-body effects beyond the KS-GGA approximation will blue shift such excitations to a region above 4 eV, where hBN excitonic effects cannot be neglected. Therefore, we disregard the $3V_N + V_B$ defect structure.

**Figure 5** depicts the electronic and optical properties of the $3V_B + V_N$ defect. We may observe that the optical spectrum presents a series of well-defined peaks from 0.59 eV to 2.38 eV. Upon inspection of the corresponding one-electron energies, we may conclude that many of these transitions involve extended occupied states near the edge of the valence band and unoccupied defect states. However, the most intense peak at 2.2 eV and the peak at 0.8 eV correspond to transitions from a defect-state resonance within the valence band to unoccupied defect states.

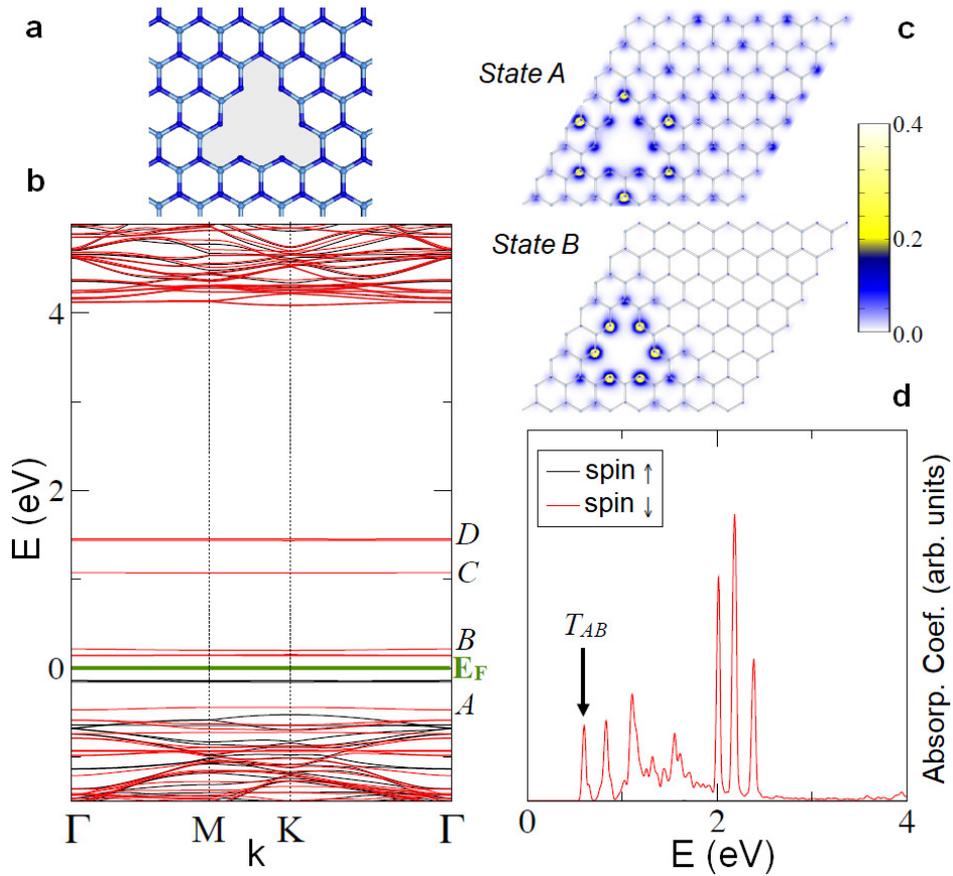

**Figure 5.** Electron and optical properties of the $3V_N + V_B$ defect. a) Geometry of the defect. b) Spin-resolved KS band structure labeling the spin-down defect levels A, B, C, D. c) Representation of the projected LDOS of state A and LUMO state B. d) Absorption spectrum corresponding to perpendicularly incident unpolarized light as obtained from the sum of the *xx* and *yy* components of the independent-particle dielectric tensor. The arrow indicates the transition at 0.59 eV between the defect states A and B.

Finally, the lowest-energy peak is due to a transition from the occupied defect state just above the valence band (state A in Figure 5) to one of the defect states of the almost degenerate LUMO level B, whereas the peak at 2 eV is likely due to a transition from the same defect state A to one of the states of the almost degenerate level D. As shown in the Supporting Information Figure S9, the details of the optical response are affected by geometrical distortions of the defect, logical consequence of the redistribution number of localized defect states within the gap. However, the passivation of all the dangling bonds of the N ions around the defect leads to the disappearance of the optical absorption below 4 eV (Figure S10). We may then conclude that one-photon emission can be centered at a $3V_B + V_N$ defect at similar but also at lower energies than the one centered at a $V_N + S_{B \rightarrow N}$ defect. Furthermore, the $3V_B + V_N$ defect and, to a lesser extent the $V_N+S_{B\rightarrow N}$ one, lead to a rich optical response in the region above 4 eV with several well-defined features involving transitions from valence band states to unoccupied defect localized states.

Although we cannot establish a direct correlation between our different experiments, or between the experiments and the theoretical calculations, our work points to the existence of several single-photon transitions associated with the different observed vacancies. Recent results point to carbon-containing defects as an important source of single photon emission in hBN, in particular the negatively charged $V_B C_N^-$ defect.[14] Our experimental data are not sensitive to the presence of impurities incorporated into the hBN lattice, so we cannot discard a partial contribution of carbon-type or other impurities in our measurements in addition to the different vacancies observed in our work and other defects not observed but that might be present.

## 3. Conclusions

We have studied few-layer hBN from thermally expanded hBN powder. From the combination of high resolution AFM images in ambient conditions, micro-photoluminescence characterization and KS-DFT calculations, we observe the existence of several single-photon transitions within the range [0.5, 4.0] eV, that we can associate to different defect structures. Our results help to understand the origin of single-photon emission in hBN, which is important to enhance the fundamental understanding of its optical and electronic properties and tuning of high quality emitters for potential future applications in fields such as quantum technologies and optoelectronics.

## 4. Methods

*Thermal expansion of hBN*: We prepared commercial hBN powder (from Sigma-Aldrich, ~1 µm, 98% purity), placing 100 mg into a quartz basket inside a tubular oven and heating up to 1000 ºC for 10 min. We then placed 10 mg of the thermally expanded hBN powder into a 20 mL vial with 10 mL of a (4:1) 2-Propanol/H$_2$O (v/v) mixture. We dispersed the sample with a shear-mixer device (IKA Ultra Turrax T25 digital) for 90 min at 25000 rpm. In the next step, in order to homogenize the sample we centrifuged the resulting suspension for 10 min at 4000 rpm. More details can be found in ref.[12].

*Atomic force microscopy*: We used an AFM from Nanotec Electronica S.L. and the WSxM software (www.wsxm.es) for both data acquisition and image processing.[55, 56] We acquired all the AFM topographical images in contact mode to avoid possible artifacts in the flake thickness measurements.[40] We employed OMCL-RC800PSA cantilevers (www.probe.olympus-global.com) with nominal spring constants of 0.05 Nm$^{-1}$ and tip radii of 15 nm. We used low forces of the order of 100 pN for imaging to ensure that the flakes would not be deformed by the tip. In order to both optimize the quality of the images and highlight the defects, the high resolution images have been both high and low pass filtered using the FFT filter and Smooth options in WSxM. Unfiltered images can be found in the Supporting Information Figure S11.

*Optical characterization*: We performed micro-photoluminescence characterization in a homemade setup. We transferred a small amount of grains consisting of a collection of thermally expanded hBN flakes to $SiO_2$/Si substrates. We mounted our samples in a liquid Helium closed loop cooled cryostat and measured at both 10 K or at room temperature. We excited the grains of hBN flakes with a linearly polarized continuous-wave excitation laser of 442, 458 and 488 nm. We focused the excitation on the sample through a 50× objective lens (NA = 0.73) to a 1.5 μm spot, and we collected the luminescence through the same lens. The pump laser was rejected using an ultra-steep long-pass filter. The light was dispersed using a single grating monochromator and recorded with a nitrogen-cooled CCD camera. Alternatively, we sent the light to an intensity interferometer (HBT setup) coupled to a lateral exit of the monochromator. We used two Perkin Helmer avalanche photodiodes connected to a PicoHarp (PicoQuant) photon counter to record single-photon events.

*Kohn-Sham density functional theory (KS-DFT) calculations*: We obtained electronic and optical properties within the framework of Kohn-Sham density functional theory (KS-DFT).[53, 54] Specifically, all the calculations are spin-polarized and use a localized basis set representation of the electron orbitals as implemented in the SIESTA code.[57, 58] Taking as reference the atomic valence configurations of B and N, we employed an optimized triple ζ single-polarized basis set. We used an auxiliary real-space grid corresponding to a plane-wave energy cutoff of 350 Ry for the evaluation of electron densities and potentials. Exchange-correlation (XC) effects are approximated by the GGA functional by Perdew, Burke, and Ernherzof,[59] whereas the electron-ion interactions are modeled with norm-conserving nonlocal Troullier-Martins pseudopotentials,[60] which include nonlinear XC corrections[61] in B and tiny ones in N to increase its transferability. The representation of the systems consists of a 6 × 6 hBN supercell accommodating a $V_N + S_{B \rightarrow N}$ defect or a 8 × 8 hBN supercell for the $3V_N + V_B$ and $3V_B + V_N$ defects, plus a vacuum spacing of 15 Å in the perpendicular direction. We performed more stringent test calculations to assess that the final results are not affected by finite-size effects. We relaxed the corresponding structures by conjugate gradient optimization until the forces were smaller than 0.04 eVÅ$^{-1}$. Finally, we evaluated the optical absorption by computing the *xx*, *yy*, and *zz* components of the imaginary part of the dielectric tensor under an independent-particle approach.

It is well known that mean-field KS-DFT may not have enough predictive accuracy to evaluate optical properties due to the systematic underestimation of band gaps and the lack of a proper treatment of electron-hole interactions in the excitation process.[62] In particular, when looking for the role of defects in the optical properties of wide band-gap materials like hBN, predicted KS-DFT transitions from valence-band to unoccupied defect states or from occupied defect states to conduction-band ones are systematically underestimated.[23] However, these shortcomings of KS-DFT prescriptions are expected to be less important when dealing with transitions involving just two defect states as long as their energy differences are much less than ~5 eV, which is the excitonic effects threshold of pristine monolayer hBN. That is, KS-DFT is a useful tool, yet not fully accurate, to identify defect geometries amenable to accommodate single-photon emission processes.[19, 25] Furthermore, unlike the computationally demanding ab-initio many body perturbation theory (MBPT),[62] KS-DFT allows the systematic investigation of how geometrical distortions or chemisorption of impurities may affect the optical properties at an affordable computational cost (see Supporting Information, where we show the sensitivity of electron and optical properties on the structural details and also the impact due to the passivation of N and B dangling bonds by hydrogen and hydroxyl groups, respectively).

**Acknowledgements**
We acknowledge financial support from the Spanish Ministry of Science and Innovation, through the "María de Maeztu" Programme for Units of Excellence in R&D (CEX2018-000805-M) and MINECO-FEDER projects MAT2016-77608-C3-1-P, MAT2016-77608-C3-3-P, PCI2018-093081, PID2019-106268GB-C31, PID2019-106268GB-C32, the "Comunidad de Madrid" grant ADITIMAT-CM (P2018/NMT-4411), the Spanish MINECO under the contract MAT2017-83722-R and the research grants PID2019-107874RB-I00 and RTI2018-099737-B-100, the European Union Seventh Framework Programme under Grant agreement No. 604391 Graphene Flagship (JTC2017/2D-Sb&Ge). This work was also supported in part by the collaborative project "Single-Photon Generation in 2D Crystals for Quantum Information" (MDM-2014-0377) funded by the Condensed Matter Physics Center (IFIMAC).

**Supporting Information**

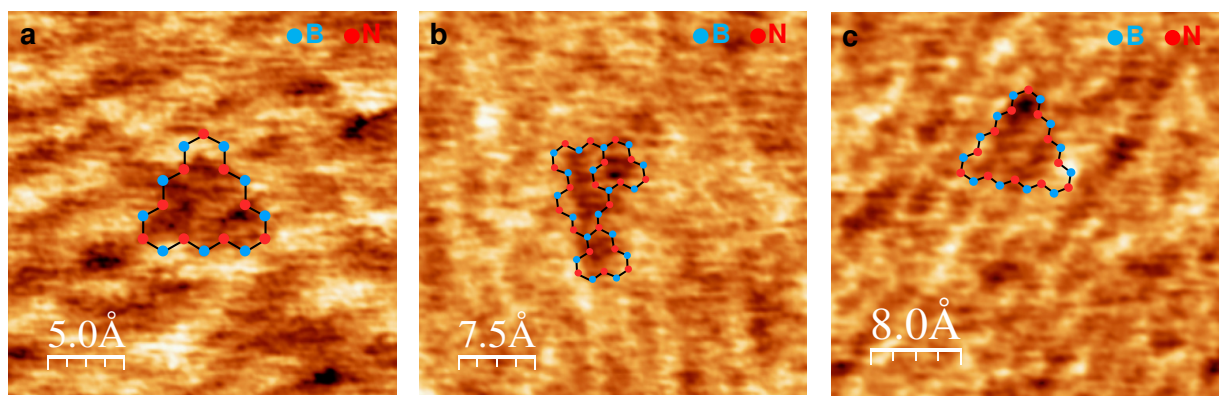

**Figure S1.** High resolution AFM topographical images showing different defects. The tentative hBN atomic structures compatible with the defects are overlaid to guide the eye. a) A $3V_B + V_N$ vacancy. b) One $V_B + V_N$, one $V_N$ and two $V_B$ vacancies. c) A $6V_B + 3V_N$ vacancy.

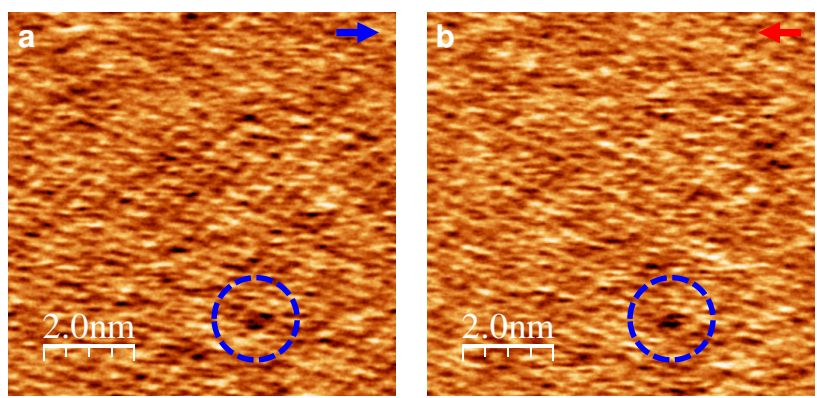

**Figure S2.** High resolution AFM topographical images of a defect in different trace/retrace scans. a) 11×11 nm² area showing the atomic lattice of hBN with an individual defect, same image as Figure 2a in the main text, reproduced here for clarity. The image corresponds to the trace scan (the tip was moving from left to right, indicated by the blue arrow in the top-right corner). b) Same defect as in a), now imaged in the retrace scan (the tip was moving from right to left, indicated by the red arrow in the top-right corner).

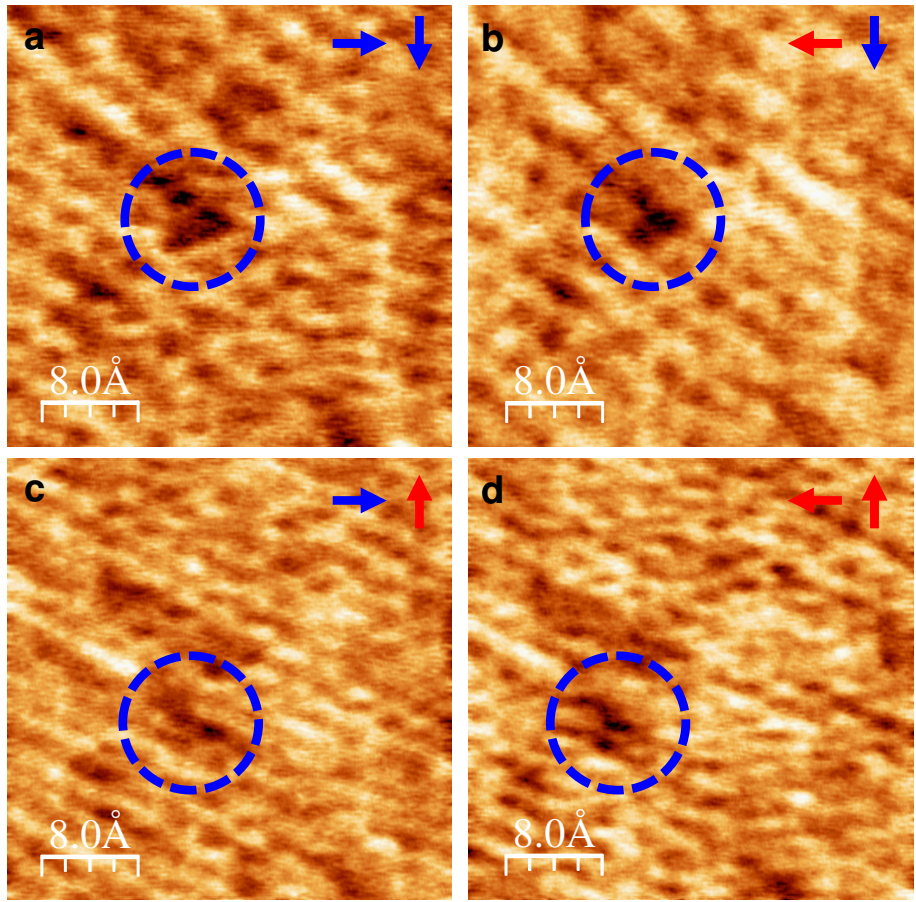

**Figure S3.** High resolution AFM topographical images of a defect in different trace/retrace and up/down scans. Same defect present in Figure 2b in the main text. Images from a) to d) correspond to the AFM tip scanning as indicated by the arrows in the top right corners of the panels: a) left-right and top-bottom, b) right-left and top-bottom, c) left-right and bottom-top, and d) right-left and bottom-top.

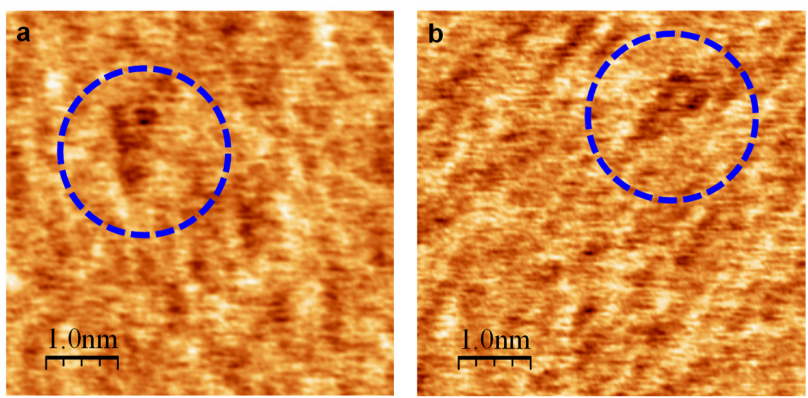

**Figure S4.** High resolution AFM topographical images showing the same defects at different scanning angles. a) Defect shown in Figure S1b at a 0° scan angle. b) Same defect at a 45° scan angle.

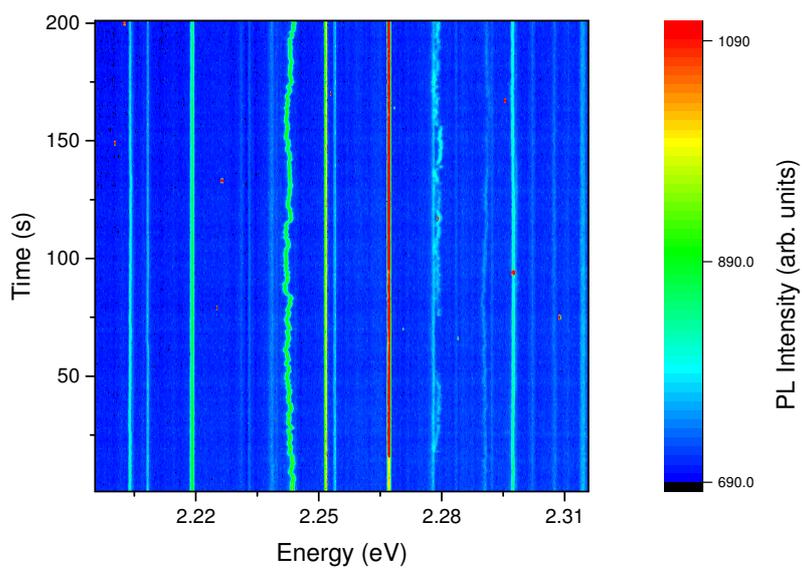

**Figure S5.** Time evolution of emission lines at 10 K, showing spectral diffusion for some of them.

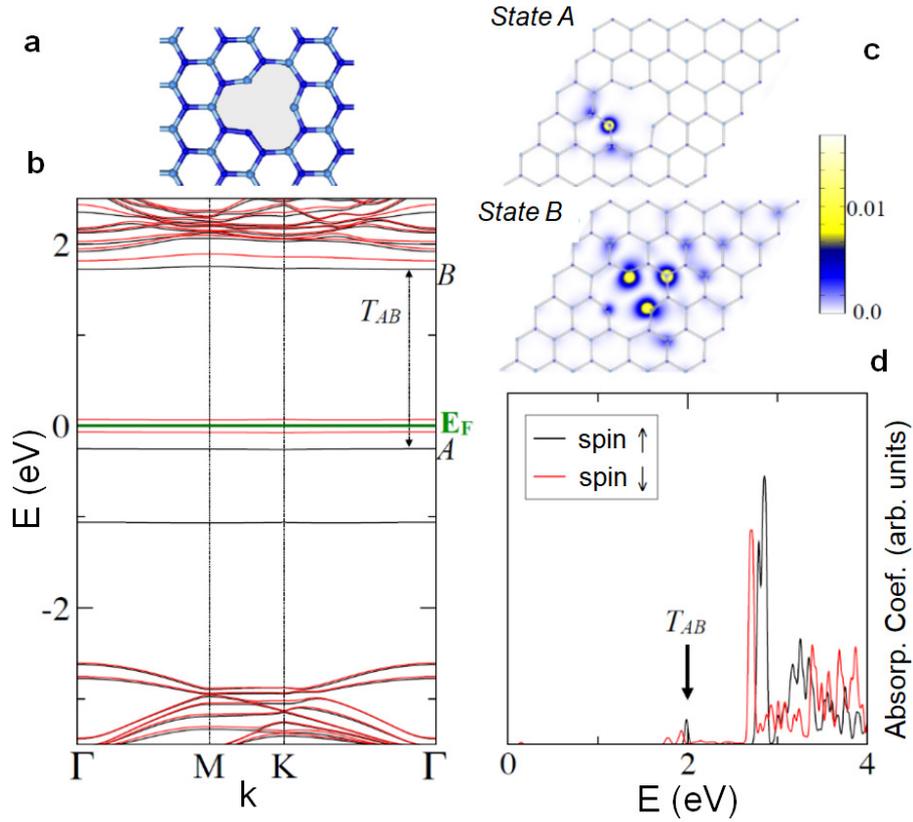

**Figure S6.** Electron and optical properties of the anti-site $V_N + S_{B\rightarrow N}$ defect (unrelaxed geometry). a) Geometry of the defect. b) Spin-resolved KS band structure. $T_{AB}$ is the optically allowed transition between the occupied defect state A and the unoccupied one B. c) Representation of the projected LDOS of states A and B. d) Absorption spectrum corresponding to perpendicularly incident unpolarized light. Note that the $T_{AB}$ transition was the dominant spectral feature upon optimization of the geometry (Figure 4 in the main text). The peak at 2.7 eV is a transition from valence-edge extended states to the LUMO defect state, and the structure around 2.9 eV is made up of transitions from the defect state 1.1 eV below the Fermi level to the state B and to conduction-band extended states. These transitions are blue shifted and lose spectral weight once the optimized geometry is considered.

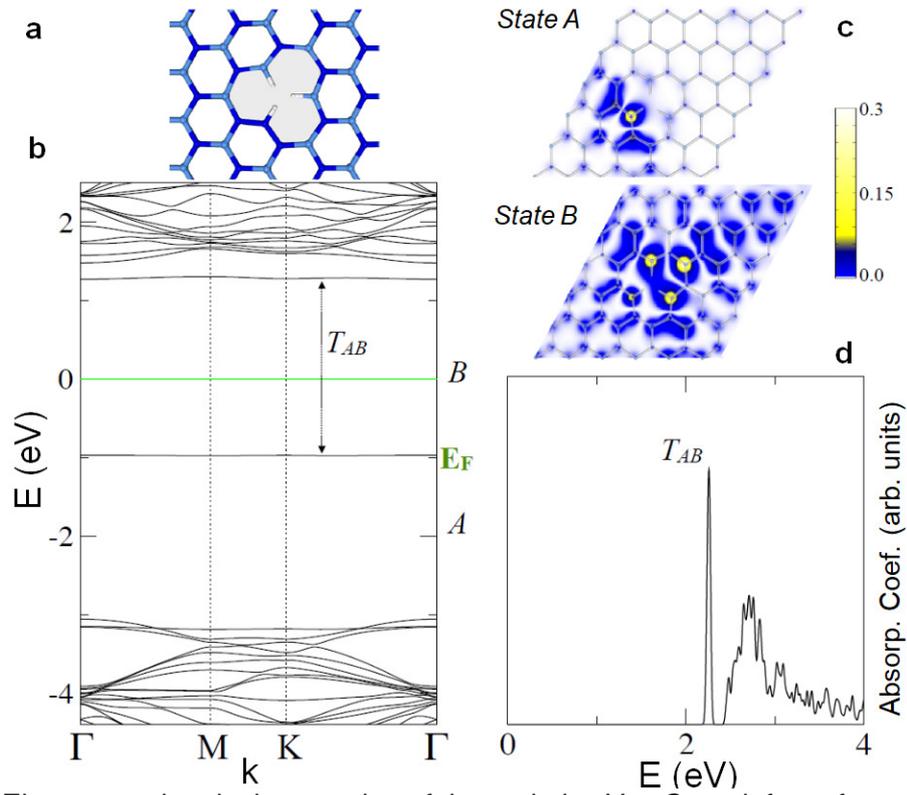

**Figure S7.** Electron and optical properties of the anti-site $V_N + S_{B \to N}$ defect after passivation of dangling bonds. a) Geometry of the defect. b) KS band structure (spin-up and spin-down levels are degenerate). $T_{AB}$ is the optically allowed transition between the occupied defect state A and the unoccupied one B. c) Representation of the projected LDOS of states A and B. d) Absorption spectrum corresponding to perpendicularly incident unpolarized light. Note that the passivation leads to a slight blue shift of the $T_{AB}$ transition.

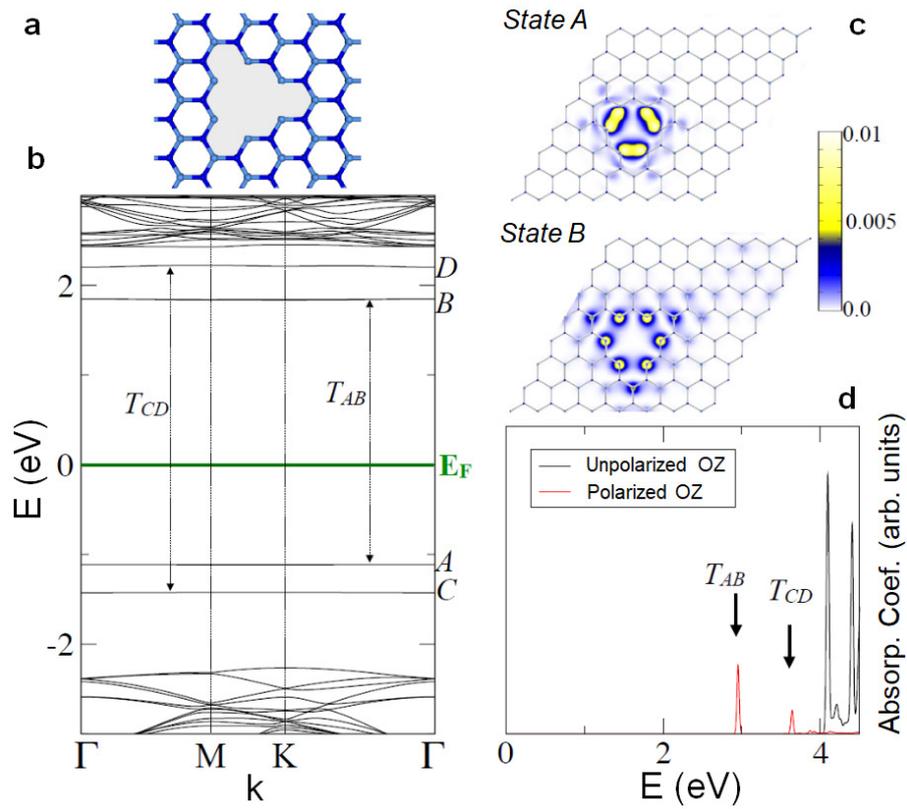

**Figure S8.** Electron and optical properties of the $3V_N + V_B$ defect. a) Geometry of the defect. b) KS band structure. $T_{AB}$ and $T_{CD}$ are optically allowed *out-of-plane* transitions between occupied defect states. c) Representation of the projected LDOS of states A (HOMO) and B (LUMO). d) Absorption spectra obtained from the sum of the *xx* and *yy* components (unpolarized OZ) and from the *zz* component (polarized OZ) of the independent-particle dielectric tensor. Note that there is no optical response below 4 eV for perpendicularly indicent light.

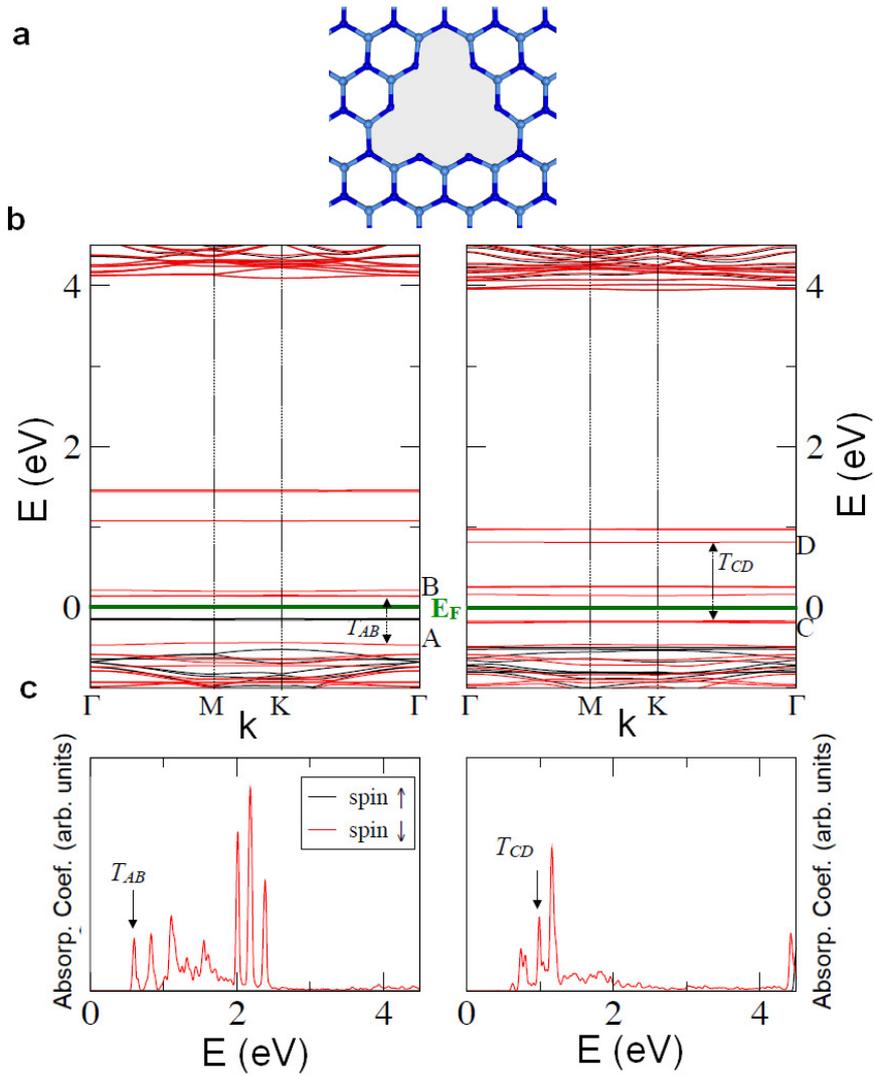

**Figure S9.** Comparison of the electron and optical properties of the $3V_B + V_N$ defect for the relaxed (left column) and the unrelaxed (right column) geometries. a) Geometry of the defect. b) Spin-resolved KS band structures. c) Absorption spectra corresponding to perpendicularly incident unpolarized light.

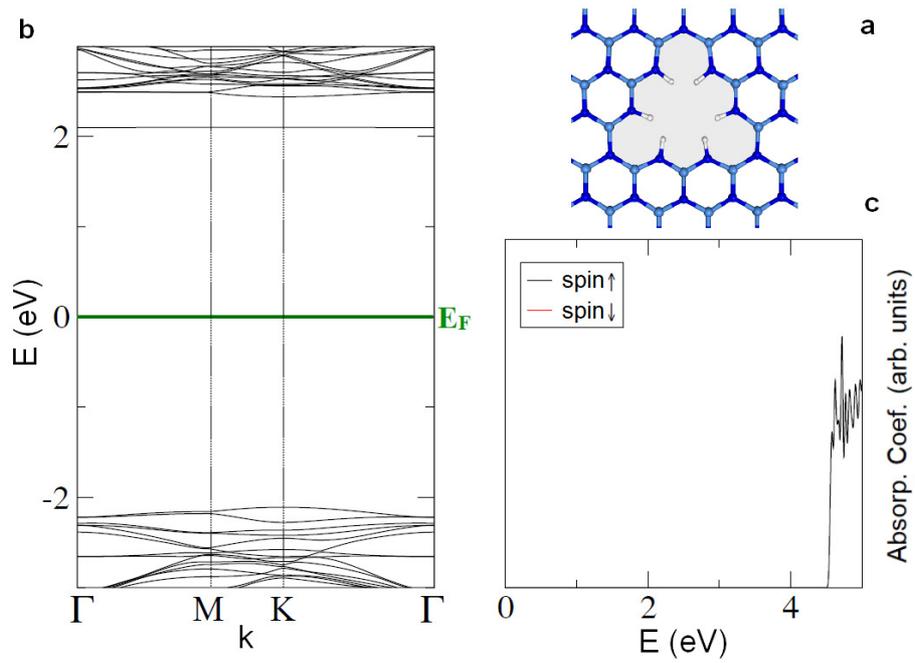

**Figure S10.** Electron and optical properties of the $3V_B + V_N$ defect after passivation of N dangling bonds. a) Geometry of the defect. b) Spin-resolved KS band structure. c) Absorption spectrum corresponding to perpendicularly incident unpolarized light. The passivation leads to the disappearance of all occupied defect states and, as a concomitant consequence, of the optical response below 4 eV.

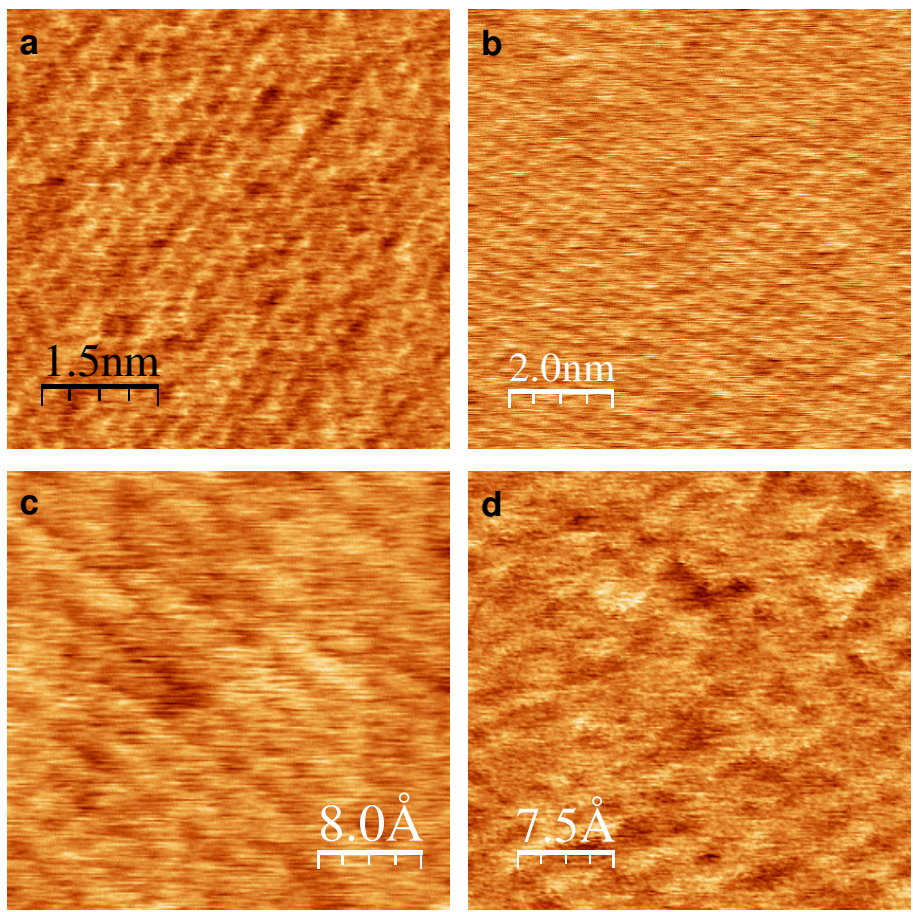

**Figure S11.** High resolution AFM unfiltered images. a) Figure 1d. b) Figure 2a. c) Unfiltered image giving rise to Figure 2b and Figure 2d. d) Unfiltered image giving rise to Figure 2c.